\documentclass[aip,reprint,nofootinbib]{revtex4-1}
\usepackage{graphicx}

\newcommand{\beq}{\begin{equation}}
\newcommand{\beqa}{\begin{eqnarray}}
\newcommand{\eeq}{\end{equation}}
\newcommand{\eeqa}{\end{eqnarray}}
\def\vec#1{\ensuremath{\mathchoice{\mbox{\boldmath$\displaystyle#1$}}
{\mbox{\boldmath$\textstyle#1$}}
{\mbox{\boldmath$\scriptstyle#1$}}
{\mbox{\boldmath$\scriptscriptstyle#1$}}}}

\begin{document}

\title{Magnetic reconnection in a comparison of topology and helicities
in two and three dimensional resistive MHD simulations}



\author{M. \v{C}emelji\'{c}}
\email[]{miki@tiara.sinica.edu.tw}
\author{R.-Y. Huang}
\altaffiliation[Also at ]{Graduate Institute of Electronic Engineering,
National
Taiwan University, Taipei 10617, Taiwan}
\affiliation{Academia Sinica, Institute of Astronomy and Astrophysics and
Theoretical Institute for Advanced Research in
Astrophysics, P.O. Box 23-141, Taipei 106, Taiwan}

\date{\today}

\begin{abstract}
Through a direct comparison between numerical simulations in two and three
dimensions, we investigate topological effects in reconnection. A simple
estimate on increase in reconnection rate in three dimensions by a
factor of $\sqrt{2}$, when compared with a two-dimensional case, is
confirmed in our simulations. We also show that both the reconnection
rate and the fraction of magnetic energy in the simulations depend
linearly on the height of the reconnection region. The degree of
structural complexity of a magnetic field and the underlying flow is
measured by current helicity and cross-helicity. We compare
results in simulations with different computational box heights.
\end{abstract}

\pacs{}

\maketitle 

\section{Introduction}
Reconnection of a magnetic field is a process in which magnetic field lines
change connection with respect to their sources. In effect, magnetic energy
is converted into kinetic and thermal energies, which accelerate and heat
the plasma. Historically, reconnection was first observed in the
solar flares and the Earth's magnetosphere, but today it is also
investigated in star formation theory and astrophysical dynamo theory.
Recently, reconnection has also been invoked in the acceleration of
cosmic rays \cite{L05}.

In solar flares, oppositely directed magnetic flux is first accumulated,
and then reconnection occurs, enabling energy transfer to kinetic energy
and heating of plasma. From such an ejection of matter, we can
observe the onset of reconnection and estimate the energy released
in this process. Recent results from measurements by the instruments
onboard the Solar Dynamic Observatory \cite{su13} have revealed new
unexpected features and show that even the morphology of the solar
reconnection is still not completely understood.

In the context of accretion disks around protostars, neutron stars and
black holes, reconnection is a part of the transport of heat, matter and
angular momentum. It enables re-arrangement of the magnetic field, after
which angular momentum can be transported from the matter that is
infalling from an accretion disk, towards the central object.
In \v{C}emelji\'{c} et al. \cite{scl1} we performed resistive
simulations of star-disk magnetospheric outflows in 2D axisymmetric
simulations. On-going reconnection is producing a fast, light
micro-ejections of matter from the close vicinity of a disk gap.
When going to three-dimensional simulations, more precise model
of reconnection is needed, as it will define the topology of
magnetic field. In the cases where flows are less ordered,
turbulent reconnection has been invoked \cite{LV99}.

The Sweet-Parker model \cite{sw58} was the first proposed model for
reconnection. Parker \cite{par57} solved time-independent,
non-ideal MHD equations for two regions of plasma with oppositely
directed magnetic fields pushed together. Particles are accelerated by
a pressure gradient, with use of the known facts about magnetic
field diffusion. Viscosity and compressibility are
assumed to be unimportant, so that the magnetic field energy converts
completely into heat. This model is robust, but gives too slow a time
for the duration of reconnection, when compared with observed data for
solar flares. Petschek \cite{pet64} proposed another model, for fast
reconnection. For energy conversion, he added stationary
slow-mode shocks between the inflow and outflow regions. This
decreased the aspect ratio of the diffusion region to the order of
unity, and increased the energy release rate, so that the observed data
were now better matched. However, his model fails in the explanation
of solar flares because fast reconnection can persist only for a very
short time period. Many aspects of the reconnection process have been
studied since, but the problem of the speed of reconnection remains
unsolved. 

Because of numerical difficulties, research on reconnection was for a
few decades constrained to two-dimensional solutions. In three
dimensional space there are more ways of reconnection than in two
dimensions, and the very nature of reconnection is different
\cite{php03}. There is still no full assessment of three dimensional
reconnection -- for a recent review, see Pontin \cite{pont11}.

Our 2D setup here is a familiar Harris current sheet, an
exact stationary solution to the problem of a current sheet separating
regions of oppositely directed magnetic field in a fully ionized
plasma \cite{har62}. It is possible to obtain a Petschek-like reconnection
in resistive-MHD simulations with uniform resistivity, but it demands
special care with the setup of boundary conditions, as described in
\cite{bat06}. To avoid this issue, we chose to set a spatially asymmetric
profile for resistivity, as suggested in \cite{bat06}. In 3D simulations,
we build a column of matter above such a Harris 2D configuration, with
resistivity dependent on height in the third direction. Because of
a modification of the resulting shocks, it enables a Petchek
reconnection also in the third dimension.

We first investigate differences between the reconnection rate in 2D and
3D numerical simulations, by comparing energies in the computational box.
Then we compare change of current helicity and cross-helicity with the
increasing height of the matter column in the third direction.

\section{Numerical setup for simulations of reconnection}
Magnetic reconnection is considered in the magnetohydrodynamic
approximation (MHD) in our simulations. There are other
possibilities. An entirely different approach, with kinetic
reconnection, is described in \cite{buch99} and references
therein. Such reconnection is a consequence of the
instability of thin current sheets in a
self-consistent consideration about ion and electron inertia, and
dissipative wave-particle resonances.

Ideal MHD, which is the study
of the dynamics of a perfectly conducting fluid, is not appropriate
for the description of reconnection. This is because a fluid must cross
the field lines of magnetic field, not follow them, as required
by ideal MHD. This means that for numerical simulations with
reconnection, a non-ideal resistive MHD is needed. For our simulations
we use the {\sc pluto} (v. 4.0) code \cite{mig12} in the parallel option.
The equations we solve are, in Gaussian cgs units:
\beqa
\frac{\partial \rho}{\partial t} + \nabla \cdot (\rho \vec{V})=0\\
\rho\left[ \frac{\partial\vec{V} }{\partial t}+ \left( \vec{V
}\cdot \nabla\right) \vec{V} \right] + \nabla p-
\frac{\vec{j}\times \vec{B}}{c} = 0 \label{mom2}\\
\frac{\partial\vec{B} }{\partial t}- \nabla \times \left( \vec{V}
\times \vec{B}-\frac{4\pi}{c}\eta \vec{j} \right)= 0 \label{faraday}\,, 
\nabla \cdot \vec{B}=0\label{faraday2}\\
\rho \left[ {\frac{\partial e}{\partial t}}
+ \left(\vec{V} \cdot \nabla\right)e \right]
+ p(\nabla \cdot\vec{V} ) -\frac{4\pi}{c^2}\eta\vec{j}^2 = 0 \,,
\label{enn}
\eeqa
with $\rho$ and p for the density and pressure, and V and B for the
velocity and magnetic field. The electric current density is given by
Amp\`{e}re's law $\vec{j}=c\nabla\times\vec{B}/{4\pi}$.
The internal energy (per unit mass) is
$e=p/(\rho(\gamma-1))$, where $\gamma$ is the
effective polytropic index. We work in the adiabatic regime, with
$\gamma=5/3$, so that the ideal gas law is $\rho=\gamma p/c_s^2$,
where $c_s$ is the sound
speed and $\eta$ is the electrical resistivity. Here we do not consider
the influence of terms other than Ohmic resistivity.

Our simulations are set in a uniform Cartesian grid with
$X\times Y \times Z$ = $(200 \times 400 \times 100)$ grid cells =
$([-1,1]\times [-2,2]\times [0,1])$ in code units. The same resolution,
equal in all three directions, is kept in simulations with different
heights and in the 2D simulations. The number of cells in the Z direction
is changed so as to obtain the same resolution with each
height of the box. Our results are compared to \cite{bat09} as a
reference. There, a nonuniform grid is used, with the smallest grid
spacing 7 times smaller in X, and 3 times smaller in the Y direction,
than in our uniform spacing. We performed simulations with different
resolutions, and found the number of cells for which the reconnection
occurred and qualitatively resembled the reference. In the results with
finer resolution, only the time needed to reach different stages varied.
With coarser resolution, reconnection did not occur, as the
current sheet was not enough resolved.

\subsection{Initial and boundary conditions}
\begin{figure}
\includegraphics[width=8cm]{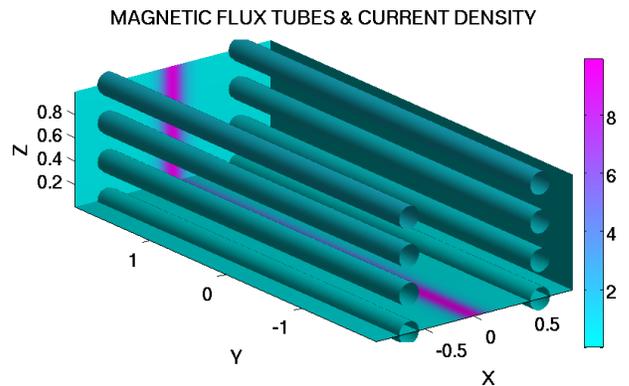}   
\caption{Setup of initial and boundary conditions in our simulations in
three dimensions. In two dimensional simulations, Z=0.
Color grading is showing a current density in code units, at the
boundary planes; the diameter of the magnetic flux tube is set
proportional to the magnetic field strength. We start with
a 2D simulation in Cartesian coordinates $X\times Y$. Increasing
the height of a box in Z direction, we compare the reconnection
rates and other interesting quantities in the flow.
}
\label{initcond3d}
\end{figure}
To ease a comparison of results, we choose the same
initial conditions for density, pressure, velocity and two regions
of oppositely directed magnetic field as in \cite{bat09}. The
so called Harris current sheet is formed initially with magnetic field
$\mathbf{B_{2}}(x)=\mathbf{y}B_{0}\tanh({\mathbf x}/b)$, which
is parallel to the Y axis and is varying in the X direction. The
amplitude of the magnetic field is chosen to be $B_0=1$, and the
initial half-width of the current sheet is $b=0.1$. We set the
plasma $\beta=0.35$, the initial pressure to $p(x)=1.25-B_2^2/2$, and
the density $\rho(x)=2p(x)/\beta$. To obtain
Petschek reconnection with spatially uniform resistivity, boundary
conditions need to be overspecified at one of the inflow boundaries
by imposing the mass density, two components of the velocity, one
component of the magnetic field, and the total energy \cite{bat06}.
To avoid such issues, we include the anomalous Ohmic resistivity
(which is for a few orders of magnitude larger than microscopic
resistivity), with a profile asymmetric in the X-Y plane and dependent
on height in the Z-direction:
\beqa
\eta=(\eta_0-\eta_1)\exp{[-(x/a_1)^2-(y/a_2)^2]}\\ \nonumber
+(\eta_0-\eta_1)\exp{-(x/a_1)^2-(z/a_3)^2}+\eta_1 \\ \nonumber
=\eta_1+(\eta_0-\eta_1)\exp{[-(x/a_1)^2]}\\ 
\cdot(\exp{[-(y/a_2)^2]} +\exp{[-(z/a_2)^2]})\nonumber \ ,
\eeqa
with the characteristic lengths of the resistivity variation in each
direction $a_1$=0.05, and $a_2$=$a_3$=0.02. Constant factors in
the resistivity $\eta_0=10^{-3}$ and $\eta_1=3\times 10^{-5}$ are
also equal to those in \cite{bat09}, for easier comparison. We
checked that $\eta_1$ is above the order of numerical resistivity at
a given resolution. The level of numerical resistivity we found by
a direct comparison of simulations with different minimum resistivity;
it is of the order of $10^{-6}$. All our simulations have been performed
with the value of the resistivity set above the numerical resistivity.
The resistivity is then set equal to $\eta$ for $y\geq 0$, or to
$\eta_0$ for $y<0$. At all the boundaries, we impose the free
(``outflow'') boundary conditions, extrapolating the flow from the
box in the ghost zones.

In the {\sc pluto} code setup we used Cartesian coordinates, ideal
equation of state, and the ``dimensional splitting'' option, which
uses Strang operator splitting to solve the equations in the
multi-dimension case. The spatial order of integration was set
as ``{\sc linear}'', meaning that a piecewise TVD linear
interpolation is applied, accurate to second order
in space. We used the second order in time Runge Kutta evolution
scheme RK2, with the Eight-Waves option for constraining
the $\nabla\cdot\vec{B}=0$ at the
truncation level. As the approximate Riemann solver, we use the
Lax-Friedrichs scheme (``tvdlf'' solver option in {\sc pluto}).

We first present the well known results in 2D, to introduce a concept of
reconnection and measurement of the reconnection rate.

\section{Results in 2D}
\begin{figure}
\hspace*{-.5cm}\includegraphics[width=9cm]{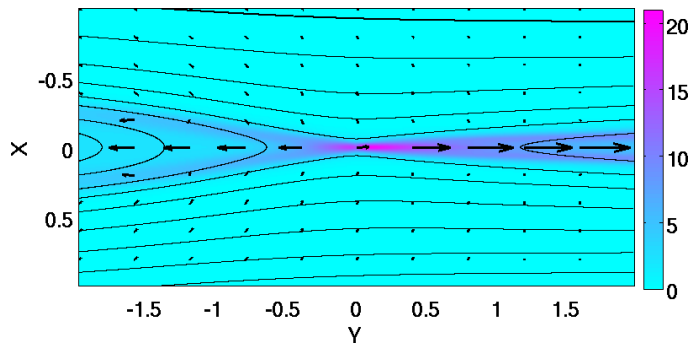}
\caption{Reconnection in two dimensions at T=30 in code units,
with current density shown in color grading, magnetic field contour
lines in solid lines, and arrows showing velocity.
}
\label{twodre}
\end{figure}
\begin{figure*}
\includegraphics[width=8cm]{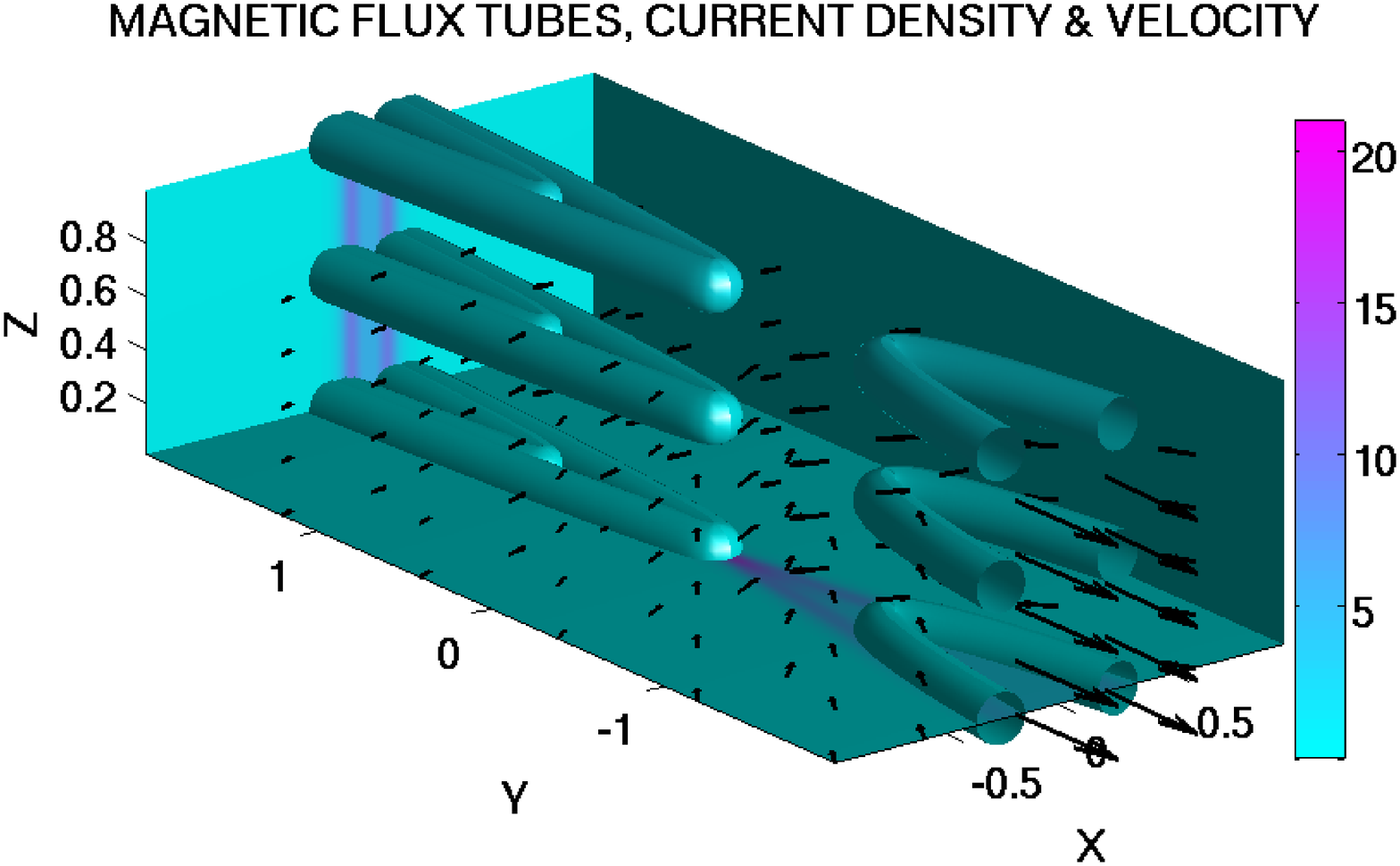}
\includegraphics[width=8cm]{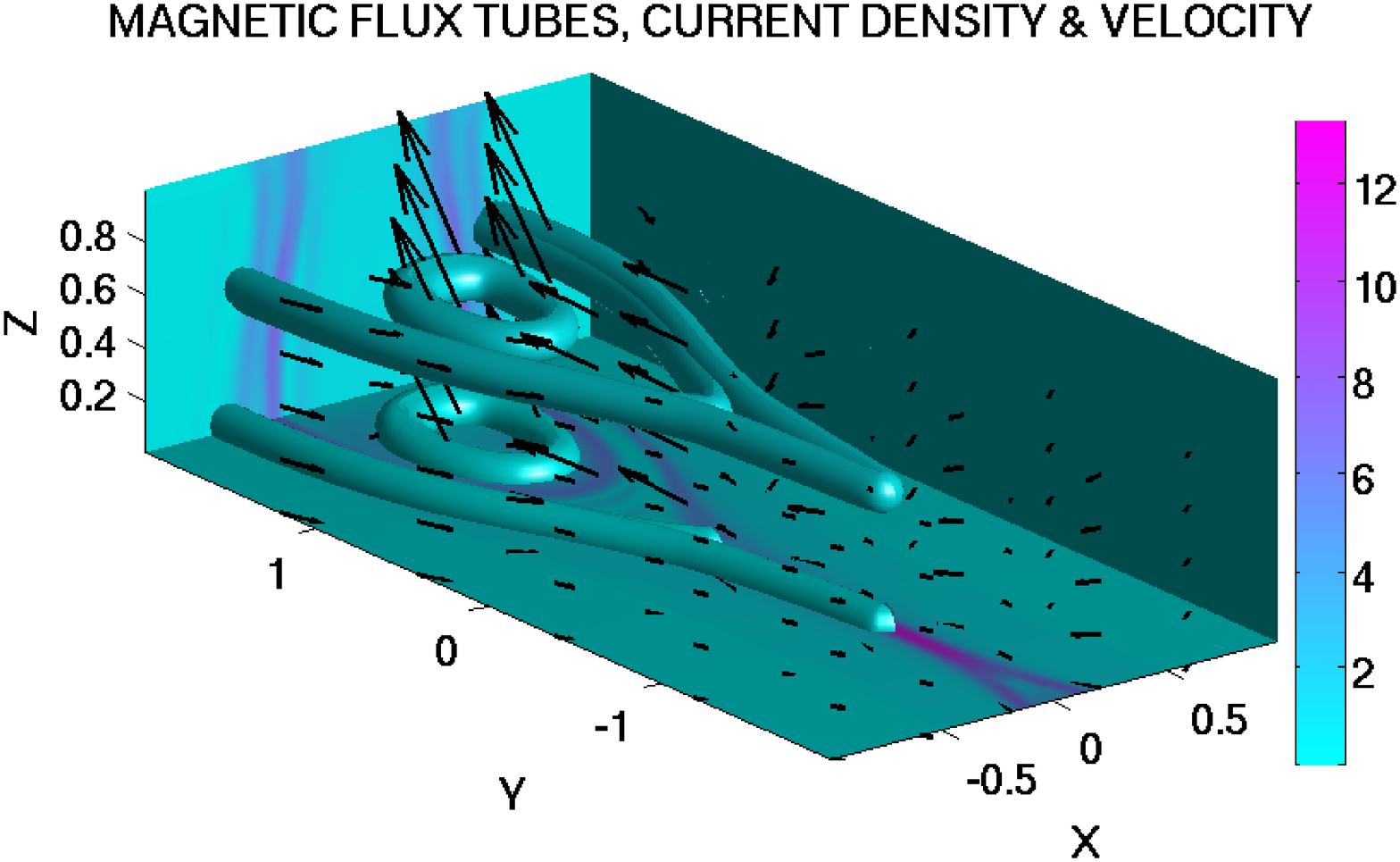}
\caption{Solutions in 3D in the first case, without the asymmetry in
resistivity in the Z direction at T=30 ({\em Left panel}), and in the
second case, with the asymmetry in the Z direction at T=70
({\em Right panel}). Color grading is showing the toroidal current
density at the boundary planes; tubes show a choice of the magnetic
flux tubes, with the diameter of the tube set proportional to the
magnetic field strength; arrows show velocity. A change in
connectivity of the magnetic flux tubes in 3D, triggered by the
asymmetry in resistivity in the vertical plane, additionally
changes, and complicates, the topology of magnetic field.
}
\label{mf3d}
\end{figure*}

Reconnection in 2D is an extensively investigated topic, but still with
inconclusive results. It is not even clear if the approach with
numerical simulations in MHD correctly describes onset of reconnection,
especially in models which invoke turbulence. Here we assume that MHD
simulations can provide correct rates.

Reconnection rate in the two-dimensional case can be
estimated in the different ways, depending on the dominating physical
properties in the system. The first such estimate was given by
Parker \cite{par57, par63}, with the Alfv\'{e}n Mach number
$M_{\mathrm {SP}}=V_{\mathrm i}/V_{\mathrm A}=S^{-1/2}$. The
Lundquist number $S=\tau_{Ohm}/\tau_{adv}=LV_{\mathrm A}/\eta $ is
the ratio of the advective over the resistive term in the Equation
(\ref{faraday}). In an astrophysical case, the length scale is so
large that the obtained reconnection rate is way too small, when
compared to the observed reconnection events on the Sun or in the
Earth's magnetosphere.

To improve the model, Petschek \cite{pet64} assumed
that the plasma can also be accelerated by slow shocks. Then the
reconnection rate is $M_{\mathrm P}=\pi/(8\ln S)$. For the current
sheet in a small region, the Petschek model gives orders of magnitude
faster reconnection than the Sweet-Parker model, up to 0.1. The
reason is that in the Petschek model the reconnection rate is not
strongly dependent on the Lundquist number. For the current sheet
extending to the whole reconnection area, both models give
the same reconnection rate.

The reconnection rate can also
be estimated from the turbulent motions \cite{LV99}. For the case
of only Ohmic resistivity, we can estimate the distance to which a
magnetic field can diffuse in time
$\tau_D$ as $\ell\sim (\eta\tau_D)^{1/2}$. This means that
two lines can merge only if their distance is of the order of
$\Delta=\ell/\sqrt{S}$. In combination with mass conservation,
one obtains the reconnection rate $M_{\mathrm {SP}}$.

In our 2D simulation, initial conditions are the same as in
Baty et al. \cite{bat09}, with asymmetric resistivity. We did not
specify fixed conditions at the inflow boundary in the X direction.
Instead, our boundary conditions are left open to ``outflow'', so
that a steady state will not be reached. The density nearby the
box boundaries along the Y direction does not change until T=70 in
code units, confining the matter inside the box. Starting from
time T=1, a Petschek reconnection is obtained. In the case shown in
Figure \ref{twodre}, the reconnection rate is of the order of
$M_{\mathrm P}=0.1$. The Figure shows a resulting velocity structure with
current density and magnetic field during reconnection, at T=30. After
T=100, because of reflections back into the computational box, and
matter leaving the box, the results become unsuitable for comparison
with steady state solutions.

\section{Results in 3D}
\begin{figure*}
\hspace*{-.6cm}\includegraphics[width=8cm,height=6cm]{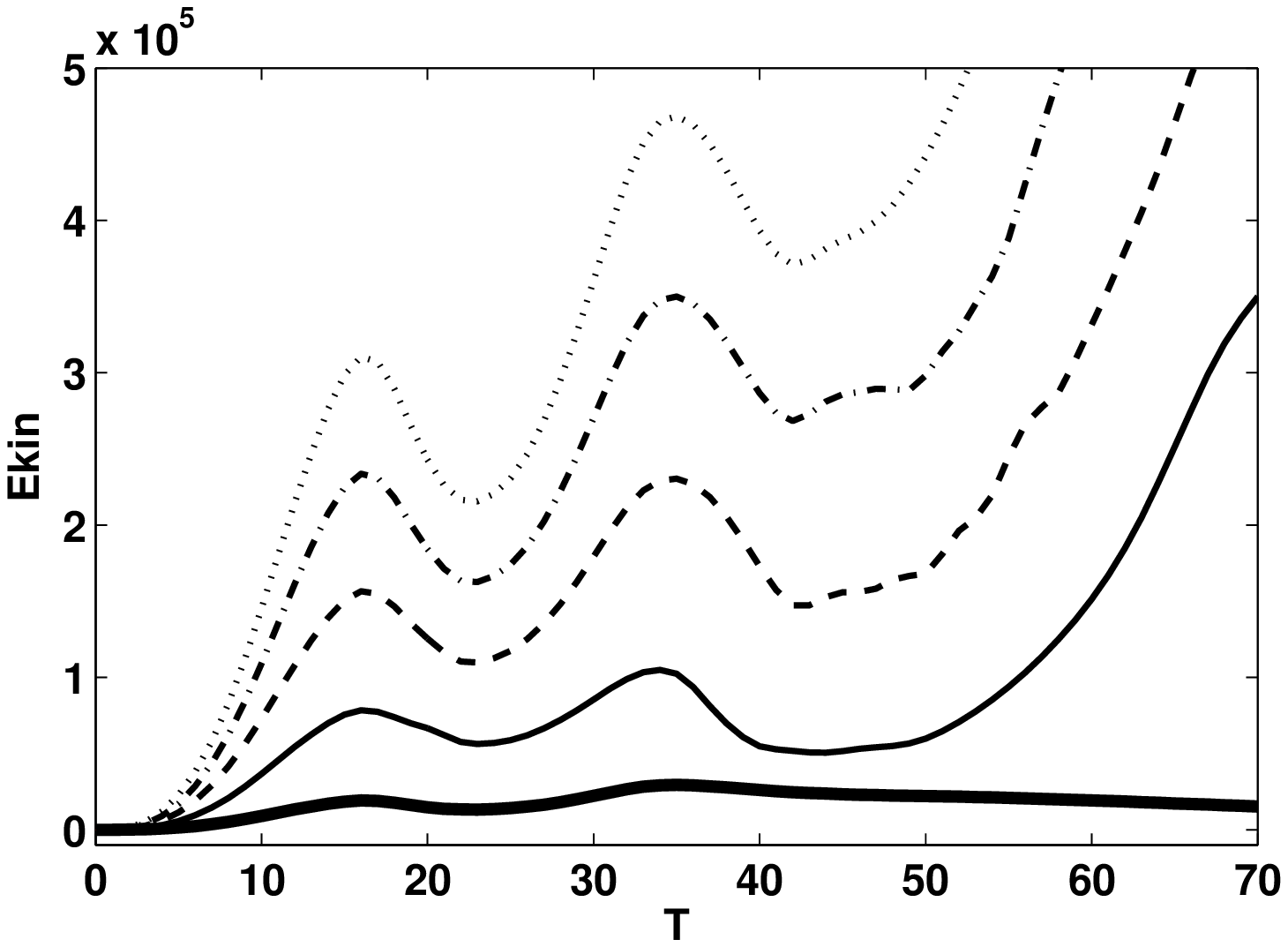}
\hspace*{-.6cm}\includegraphics[width=8cm,height=6cm]{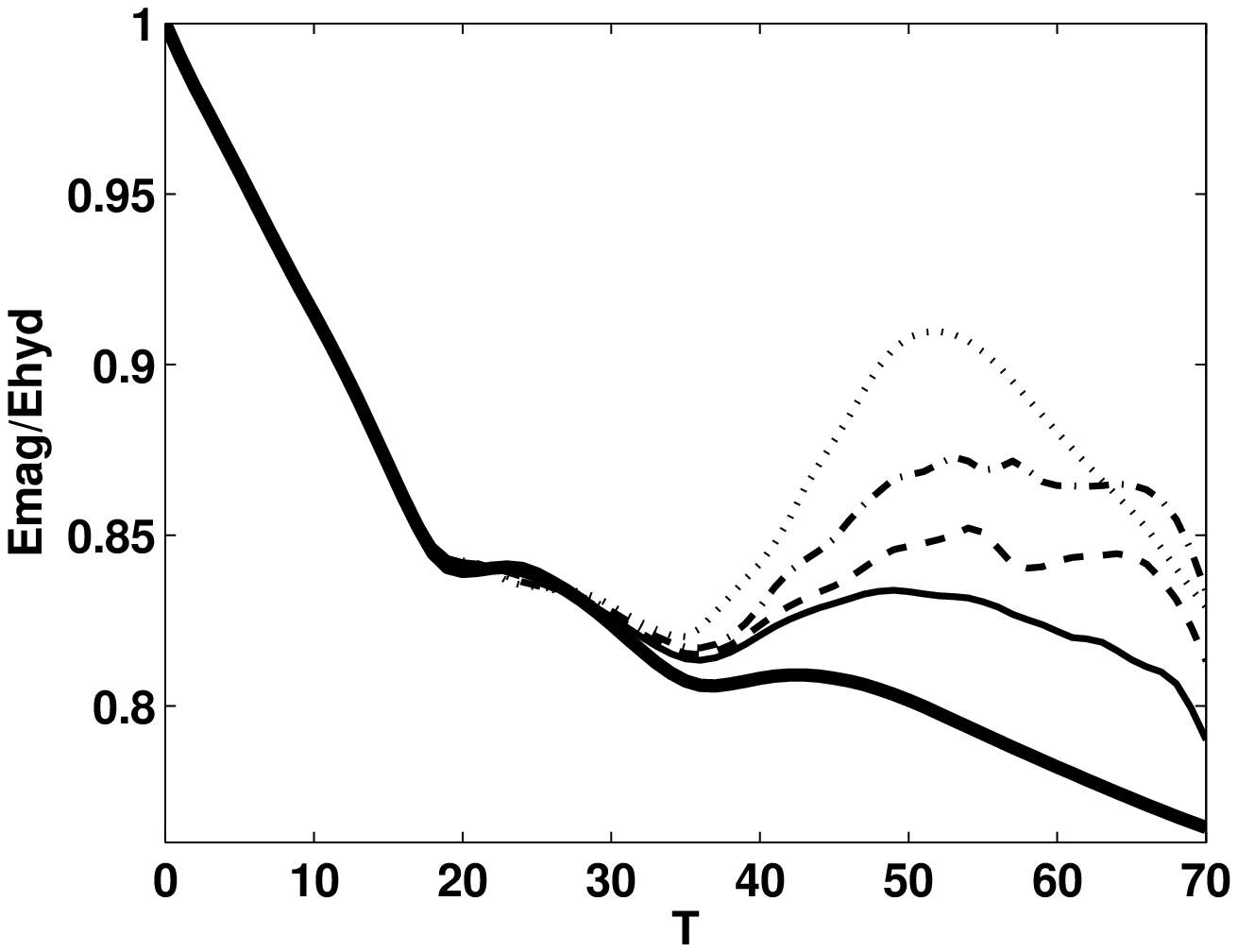} 
\caption{Time dependence of the energy with different heights of the
computational box in our simulations with reconnection in all three
directions. In the {\em Left panel} is shown the kinetic energy, and
in the {\em Right panel} is shown the ratio of magnetic to 
the sum of internal and kinetic energy. Results with different heights
of the box h=1, 2, 3 and 4, are shown in solid, dashed, dot-dashed and
dotted lines, respectively. In thick solid line is shown the result with
height h=0.25 in the case with reconnection only in the X-Y plane,
which is our reference 2D case. Kinetic energy during
the build-up of reconnection is linearly increasing with height of the
computational box, with the factor of proportionality about 2. The
fraction of magnetic energy is steadily decreasing with time, until
the reconnection in the third direction starts; then it increases
proportionally with height.
}
\label{enratz}
\end{figure*}
\begin{figure}
\hspace*{-.6cm}\includegraphics[width=8cm,height=6cm]{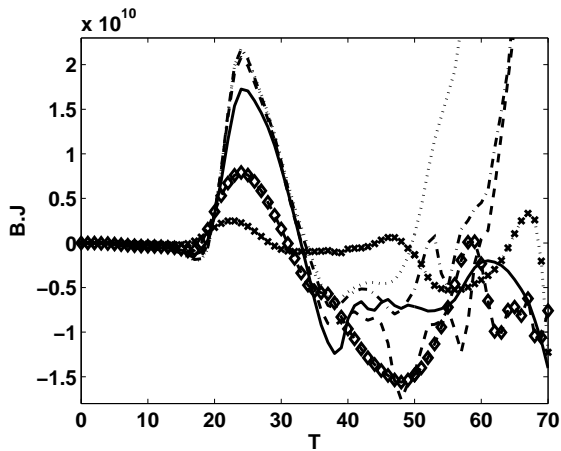}
\caption{Time evolution of current helicity in $X\leq 0$ half of the
computational box. Results with different heights of the box
h=0.25, 0.5, 1, 2, 3 and 4, are shown in black cross and diamond
marked, solid, dashed, dot-dashed and dotted lines, respectively.
}
\label{hcplot}
\end{figure}
\begin{figure}
\hspace*{-.6cm}\includegraphics[width=8cm,height=6cm]{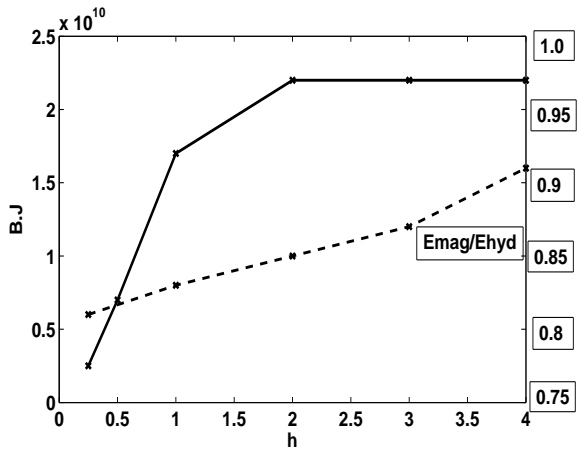}
\caption{In solid line is shown maximum variation in current helicity with
height of the computational box, during the increase in reconnection.
Values are taken from results at T=23, shown in Figure \ref{hcplot}. In
dashed line is shown the maximum variation of fraction of magnetic energy
in dependence of h, taken from results about T=50 shown in
Figure \ref{enratz}. Note the different scales in the vertical axes of
the plot; the right side axis denotes the fraction of magnetic energy.
}
\label{hbjgr}
\end{figure} 
\begin{figure}
\hspace*{-.6cm}\includegraphics[width=8cm,height=6cm]{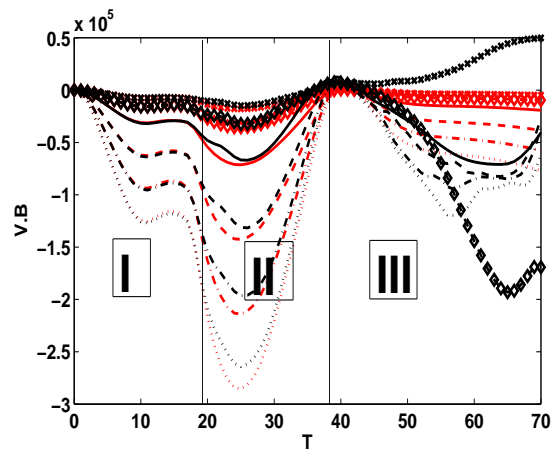}
\caption{Time evolution of cross helicity in $X\leq 0$ half of the
computational box. Lines have the same meaning as in Figure
\ref{hcplot}. Results in the first case, with reconnection only 
in the X-Y plane, are shown in red lines, with the same line
style coding. Three time intervals mark time of build-up of
reconnection in the X-Y plane (I), increase in reconnection
in the Z-direction (II), and short increase in the fraction of magnetic
energy, during the re-organization of the magnetic field because of
reconnection (III). 
}
\label{hvbplot}
\end{figure}
\begin{figure}
\hspace*{-.6cm}\includegraphics[width=8cm,height=6cm]{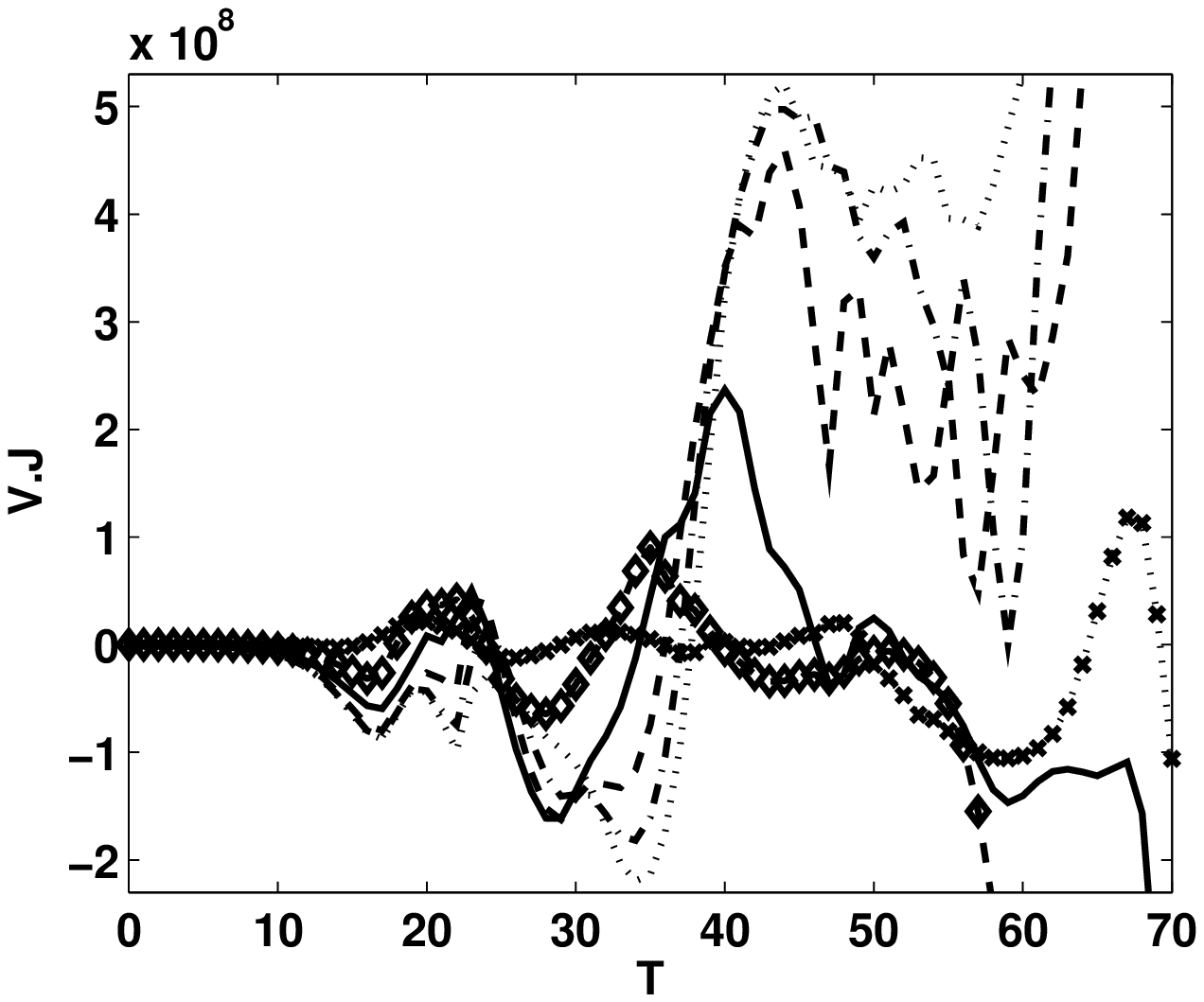}
\caption{Time evolution of ``mixed helicity'' in $X\leq 0$ half of the
computational box. Lines have the same meaning as in Figure \ref{hcplot}.
}
\label{hvjplot}
\end{figure} 

When we set up the 3D simulation with the same asymmetry in resistivity
as in the 2D case, and add the dependence of the resistivity with Z,
we obtain in each X-Y parallel plane results similar to the 2D case, as
shown in Figure \ref{mf3d}. This is because of the current sheet at
X=0 along the Y-Z plane. Without some perturbation, which would break
the stability of this current sheet, a flow does not really have
three, but still only two degrees of freedom, as if the 2D simulations
would be stacked atop each other. Instead of introducing the
perturbation -- as has been done for example in simulations with Hall
resistivity by Huba \& Rudakov \cite{HR02} -- we rather introduce
the dependence of resistivity with height (the Z direction) in our
prescription of resistivity, in the same way as we did in the X-Y plane.
For small heights of the box, the result still resembles the 2D result.
With increase in height, shocks are modified in the Z-direction,
so that at each height, the shock is of different density. This enables
the Petschek reconnection in the Y-Z plane, which is expelling matter
in the Z-direction. In Figure \ref{mf3d} both reconnection in the
X-Y and Y-Z plane are visible.

In Priest \& Schrijver \cite{ps99} it has been estimated that one additional
degree of freedom results in an increase by a factor of the order
of $\sqrt{2}$ in the reconnection rate. To estimate the reconnection
rate in 3D, we compute the energy in the computational box. For the
Lundquist number S, from the Sweet-Parker reconnection rate it
follows that
$V_{\mathrm {i,3D}}^2/V_{\mathrm {i,2D}}^2=E_{\mathrm {k,3D}}/E_{\mathrm
{k,2D}}=2$. This means that if in 3D simulations the reconnection
rate increases for $\sqrt{2}$, compared to the 2D simulation, the
corresponding kinetic energy increases by a factor of 2.
We can now verify if this estimate is true.

The results for time dependence of the integral kinetic energy
for the different heights of the box are shown in Figure \ref{enratz}.
As in the 2D case, we compare results during the phase of the simulation
when reconnection rate increases, and boundaries still do not change
much from the initial conditions. We find agreement with the above
prediction: that kinetic energy in a 3D simulation is about double
the energy in 2D, for the same length scale. We also find that the
trend continues linearly with increase in height of the box.
After the peak in reconnection (about T=50 in our simulations with
reconnection in both the X-Y and Y-Z planes), as was the case in 2D
simulations, reflections back into the computational box and matter
leaving the box make results for different heights incomparable. With less
density, matter leaves the box with higher velocity, and the kinetic energy
increases. For maintaining the stationary reconnection rate, driving at the
boundaries would be needed. Here we do not investigate such a case.

In Aly \cite{aly84,aly91} it is argued that the maximum magnetic energy
the force-free system can store, is in the configuration
without loops in the magnetic field. Any loop would, when straightened,
lead to an ``open field'' of larger energy. Our simulations are not in
the force-free regime, but measurement of the dependence of this ratio
with decrease in height of the computational box could still be
informative. Our result is shown in Figure \ref{enratz}. The
fraction of magnetic energy in total energy increases linearly with
the increase in box height.

A general measure of reconnection should anticipate not only well
ordered, but also complex reconnection, for example in the case of
reconnection in turbulent fluids. A change in topology of magnetic field
can to some degree be measured by a ``knottedness of tangled vortex
lines'' \cite{mof69}. For any vector field $\vec{F}$, the quantity
$\vec{F}\cdot(\nabla\times\vec{F})$ is called the helicity density. The
integral over a closed volume is then the helicity. For a vector field
which is invariant to the change from a right-hand to left-hand
orientation of the coordinates, helicity is zero \cite{mof78}. One such
example are our simulations here, which have a mirror symmetry
across the Y-Z plane at X=0. To obtain non-zero results, we always
integrate only over half of the computational box, in the interval
($-0.5\leq x\leq 0$). In a non-symmetric case, this integral should be
taken across the whole computational box.

We compute three quantities related to helicity, to measure the
degree of complexity of the velocity and magnetic
field: current helicity H$_{\mathrm c}$ and cross-helicity
H$_{\mathrm {VB}}$ 
\beqa
H_{\mathrm c}=\int\vec{B}\cdot(\nabla\times\vec{B})dV,\ H_{\mathrm {VB}}=\int\vec{V}\cdot\vec{B}dV\ ,
\eeqa
for which results are shown in Figures \ref{hcplot} and \ref{hvbplot}, 
and a quantity H$_{\mathrm {VJ}}$ which we provisionally
call ``mixed helicity'' 
\beq
H_{\mathrm {VJ}}=\int\vec{V}\cdot\vec{J}dV,
\eeq
shown in Figure \ref{hvjplot}. Integrations are always performed over
half of the computational box.

There are three separate time intervals, marked in Figure
\ref{hvbplot}, which are present in all the results: (I) build-up of
reconnection in the X-Y plane, (II) increase in reconnection
in the Z-direction, and (III) short increase in the fraction of
magnetic energy, during the re-organization of the magnetic field
because of reconnection.

Increase in current helicity with height of the computational box during
the interval II is shown in Figure \ref{hbjgr}, together with the
change in energy. For more than double
height of the box, increase in H$_{\mathrm c}$ is small during the
interval II. Only during the interval III, when reconnection in the
third direction is fully established, current helicity increases.
Cross helicity is better describing the three dimensional reconnection,
as it increases linearly with height of the box for the full 3D
reconnection, with h$\ge 1$. For smaller heights, h=0.25 and 0.5,
reconnection in the Z direction does not follow the trend -- this
could be an artifact of our setup, with directional asymmetry in
resistivity. Cross helicity, shown in Figure \ref{hvbplot} is
also clearly showing difference between reconnection in 2D and in
3D throughout both intervals II and III.

``Mixed helicity'', shown in Figure \ref{hvjplot} shows there is more
structure to fields of velocity and current in all the intervals from
I-III, than revealed in the other two helicities or energies. It
remains to be seen, in simulations with less ordered, or turbulent
reconnection, how much of this structure is related to strictly
directional asymmetry in our setup of resistivity.

\section{Summary}
We have presented new results with the direct comparison of numerical
simulations of reconnection in two and three dimensions. Reconnection
in our simulations is facilitated by an asymmetry in the Ohmic
resistivity. Without asymmetry, reconnection does not occur in
our setup. Asymmetry in the X-Y plane is enabling the reconnection in that
plane, and dependence of the resistivity with height in Z-direction is
changing the shocks in the Z-direction, so that the Petschek reconnection
starts also in the Y-Z plane.

By comparing the integral kinetic energy in 2D and 3D computations, we
find that the 3D simulation proceeds with a reconnection rate which is
for a factor $\sqrt{2}$ larger from the rate in the 2D simulation. This
finding confirms the simple analytic estimate from Priest \& Schrijver
\cite{ps99}. We also show that a fraction of magnetic energy in total
energy is increasing linearly with the increase in box height.

We obtained our results in the case when reconnection was set by an
asymmetry in resistivity. There are other means of facilitating
reconnection. One natural generalization from a 2D simulation of
X-point collapse of a magnetic field into a localized current layer
in a 3D situation is to obtain points in space at which the magnetic
field strength is zero -- 3D null points. Topology of such points is
characterized by a pair of field lines forming a separatrix surface,
which separates portions of magnetic field which are of different
topologies. Yet another way to form a current sheet in 3D is to
connect two such null points -- forming a separator line
(\cite{pont11} and references therein). Reconnection in 3D is also
possible without null points, in regions in which field lines are
non-trivially linked with each other (as for example in braided
magnetic fields or as the result of some ideal instability). Among
others, there is also a possibility of a current sheet formation by
a motion of a magnetic field line footpoint \cite{par72}.

Comparison of results in the various approaches mentioned above is not
straightforward; this is why we decided for more general measures.
By computing current helicity, cross helicity and ``mixed helicity''
in our choice of setup, we find three characteristic time
intervals in all our simulations. In two of them, reconnection in the
three dimensional simulation increasingly differs from the
corresponding reconnection in the two dimensional simulation, and the
results also depend on the height of the reconnection region.

It remains to be studied if reconnection in three dimensional
simulations is well described by energies and helicities in the
cases of less ordered, and of turbulent reconnection. In a future
study we will also include other resistive terms, and apply the
results in models of resistivity in simulations of reconnection in
astrophysical outflows.

\begin{acknowledgments}
We thank A. Mignone and his team of contributors for the possibility to
use the {\sc pluto} code. M.\v{C}. thanks R. Krasnopolsky for helpful
discussions.
\end{acknowledgments}

\end{document}